\documentclass[11pt,a4paper]{article}
\usepackage{amsmath, bbm, longtable, paralist, setspace}
\usepackage[top=1.56in, bottom=1.56in, left=1.2in, right=1.2in]{geometry}

\usepackage[font=small]{caption}
\DeclareCaptionStyle{italic}[justification=centering]{labelfont={bf},textfont={it},labelsep=colon}
\usepackage{graphicx,psfrag,epsf}
\usepackage{enumerate}
\usepackage{natbib}
\usepackage{url} 
\usepackage{booktabs, subfig, bm, paralist, mathpazo, tikz, microtype, dsfont, rotating} 
\usepackage[pdftex,colorlinks=true]{hyperref}
\definecolor{darkblue}{rgb}{0,0,.6}
\hypersetup{citecolor=darkblue,linkcolor=darkblue,urlcolor=darkblue}
\newcommand{\X}{\mathcal{X}}

\newcommand{\blind}{0}

\newcommand{\Z}{\mathcal{Z}}
\newcommand{\Y}{\mathcal{Y}}
\newcommand{\W}{\mathcal{W}}
\newcommand{\EE}{\mathcal{E}}
\renewcommand{\P}{\text{P}}
\renewcommand{\H}{\mathcal{H}}
\graphicspath{{plots/}}

\newsavebox\CBox
\def\textBF#1{\sbox\CBox{#1}\resizebox{\wd\CBox}{\ht\CBox}{\textbf{#1}}}

\definecolor{a0}{rgb}{0.0, 0.5, 0.0}
\definecolor{bistre}{rgb}{0.24, 0.17, 0.12}
\definecolor{amethyst}{rgb}{0.6, 0.4, 0.8}
\definecolor{blue-violet}{rgb}{0.54, 0.17, 0.89}
\definecolor{Rcolor}{RGB}{150,160,190}
\definecolor{blush}{rgb}{0.87, 0.36, 0.51}
\definecolor{brightturquoise}{rgb}{0.03, 0.91, 0.87}
\definecolor{burntorange}{rgb}{0.8, 0.33, 0.0}

\usepackage{orcidlink}
\newcommand{\Rlogo}{\protect\includegraphics[height=1.8ex,keepaspectratio]{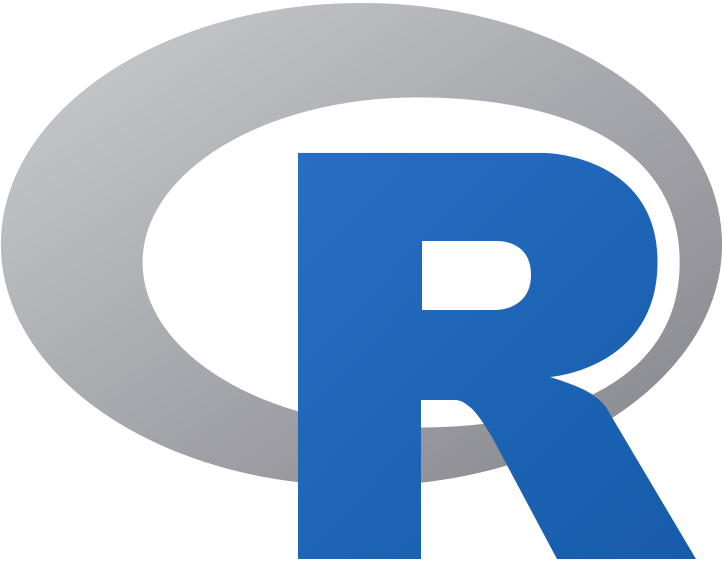}}

\date{}
\AtBeginDocument{}

\begin{document}

\def\spacingset#1{\renewcommand{\baselinestretch}
{#1}\small\normalsize} \spacingset{1}

\if0\blind
{
  \title{\bf Bootstrap prediction intervals for the age distribution of life-table death counts}
  \author{\normalsize Han Lin Shang\orcidlink{0000-0003-1769-6430}\thanks{Postal address: Department of Actuarial Studies and Business Analytics, Level 7, 4 Eastern Rd, Macquarie University, Sydney, NSW 2109, Australia; Telephone: +61(2) 9850 4689; Email: hanlin.shang@mq.edu.au}  \hspace{.2cm}\\
    \normalsize Department of Actuarial Studies and Business Analytics \\
    \normalsize Macquarie University
}
  \maketitle
} \fi

\if1\blind
{
\title{\bf Bootstrap prediction intervals for the age distribution of life-table deaths}
  \maketitle
} \fi

\bigskip

\begin{abstract}
We introduce a nonparametric bootstrap procedure based on a dynamic factor model to construct pointwise prediction intervals for period life-table death counts. The age distribution of death counts is an example of constrained data, which are nonnegative and have a constrained integral. A centered log-ratio transformation is used to remove the constraints. With a time series of unconstrained data, we introduce our bootstrap method to construct prediction intervals, thereby quantifying forecast uncertainty. The bootstrap method utilizes a dynamic factor model to capture both nonstationary and stationary patterns through a two-stage functional principal component analysis. To capture parameter uncertainty, the estimated principal component scores and model residuals are sampled with replacement. Using the age- and sex-specific life-table deaths for Australia and the United Kingdom, we study the empirical coverage probabilities and compare them with the nominal ones. The bootstrap method has superior interval forecast accuracy, especially for the one-step-ahead forecast horizon.

\vspace{.1in}
\noindent Keywords: compositional data analysis; functional principal component analysis; functional time series analysis; kernel sandwich estimator; long-run covariance function
\end{abstract}

\spacingset{1.58}

\newpage
\section{Introduction}\label{sec:1}

Actuaries and demographers have long been interested in developing mortality modeling and forecasting methods for pricing and government planning. In the literature on human mortality, three functions are widely studied: mortality rate, survival function, and life-table deaths. Although these three functions are complementary \citep[see, e.g.,][]{PHG01, DMW09}, they differ by the amount of constraints. Most literature has been focused on developing new approaches for forecasting age-specific mortality rates \citep[see, e.g.,][for comprehensive reviews]{Booth06, BT08, BCB23}. Instead of modeling central mortality rates, we consider modeling the life-table deaths as an example of probability density function \citep[see, e.g.,][]{BKC20}. Observed over a period, we can model and forecast a redistribution of life-table deaths, where deaths at younger ages are gradually shifted toward older ages due to longevity. Apart from providing an informative description of the mortality experience of a population, the life-table deaths yield readily available information on `central longevity indicators' \citep[see, e.g.,][]{CRT+05, CanudasRomo10} and lifespan variability \citep[see, e.g.,][]{Robine01, VZV11, VC13, AV18, AVB+20}.

To model the age distribution of deaths, we resort to compositional data analysis (CoDa), which is important in many scientific fields, such as actuarial science, demography, and statistics. In actuarial science, \cite{SH20} and \cite{SHX22} apply the centered log-ratio (clr) transformation of \cite{Aitchison86} to model and forecast life-table deaths and compute lifetime or fixed-term annuity. In demography, \cite{Oeppen08} and \cite{BCO+17, BSO+18} apply the clr transformation to obtain unconstrained data, which can be modeled via principal component analysis. In statistics, \cite{SM22} apply the CoDa to model cause-specific mortality data. \cite{Delicado11} apply the CoDa to analyze density functions over space, while \cite{KMP+19} model and forecast a financial time series of density functions.

In the aforementioned literature, a missing gap is in the construction of prediction intervals for the age distribution of deaths. However, prediction intervals are valuable for assessing the probabilistic uncertainty associated with point forecasts. As was emphasized by \cite{Chatfield93} and \cite{Chatfield00}, it is essential to provide interval forecasts as well as point forecasts to
\begin{inparaenum}
\item[1)] assess future uncertainty levels;
\item[2)] enable different strategies to be planned for a range of possible outcomes indicated by the interval forecasts;
\item[3)] compare forecasts from different methods more thoroughly and
\item[4)] explore different scenarios based on various assumptions.
\end{inparaenum}

To construct prediction intervals with accurate coverage, it is essential to consider all sources of forecast uncertainty. The forecast uncertainty stems from both systematic deviations (e.g., due to parameter uncertainty) and random fluctuations (e.g., due to model error).

We propose a nonparametric bootstrap procedure based on the dynamic factor model (DFM) of \cite{MGG22} to generate bootstrap forecasts of life-table deaths in Section~\ref{sec:3}. The unconstrained data are treated as nonstationary functional time series through the clr transformation. We apply the DFM to decompose the original series into separable nonstationary and stationary parts. The nonstationary part is modeled by taking the first-order differencing of the original series and then decomposing it onto a set of functional principal components. By projecting these components onto the original series, we obtain the associated principal component scores. As a byproduct, we also obtain the model residual functions, which may still exhibit temporal dependence after implementing an independent and identically distributed (i.i.d.) test. Thus, a second-stage functional principal component decomposition is required for modeling any remaining temporal dependence exhibited in the residual functions.

Using the age- and sex-specific life-table deaths in Australia from 1921 to 2020 and the United Kingdom (UK) from 1922 to 2021 in Section~\ref{sec:2}, we study the interval forecast accuracy of the bootstrap method in Section~\ref{sec:4}. The conclusion is presented in Section~\ref{sec:5}, along with some ideas on how the methodology can be further extended.

\section{Period life-table deaths}\label{sec:2}

We consider age- and sex-specific life-table deaths in Australia and the UK. As developed countries, these two nations were selected due to a relatively higher data quality, and the data are obtained from the \cite{HMD24}. We study life-table deaths with a radix of~$100,000$. For a given year $t$, the population experienced 100,000 births at age 0. As age increases, the number of survivors decreases gradually to zero at the last age group. There are 111 ages, which are $0$, $1$, $\dots$, $109$, and $110$ and above. Due to rounding, there are zero counts for age 110+ at some years. To solve this issue, we work with the probability of dying (i.e., $q_x$) and the life-table radix to recalculate our estimated life-table deaths (up to six decimal places). We obtain more detailed and smoothed death counts than those reported in \cite{HMD24}.

\begin{figure}[!htb]
\centering
\includegraphics[width=8.42cm]{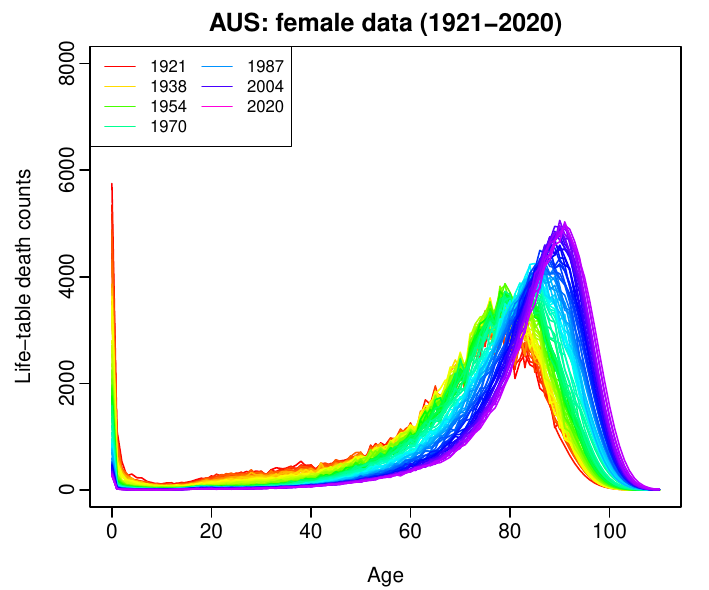}
\quad
\includegraphics[width=8.42cm]{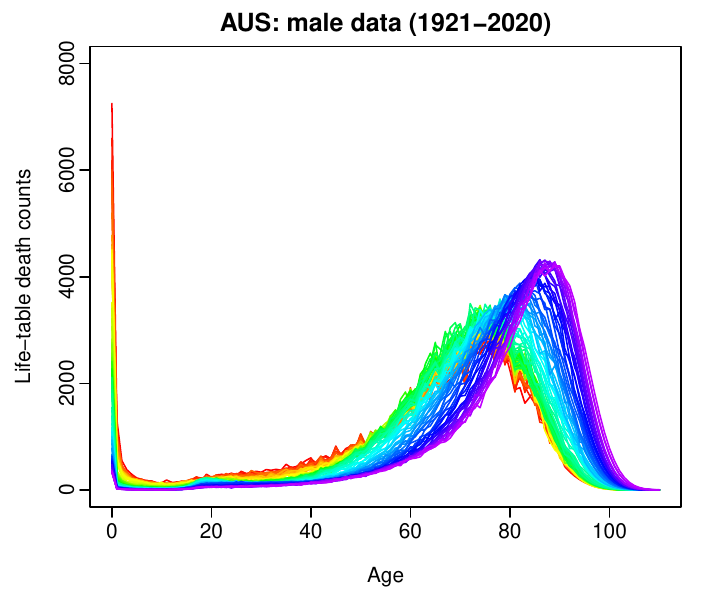}
\caption{\small Rainbow plots of age- and sex-specific life-table deaths from 1921 to 2020 in a single-year group. The oldest years are shown in red, with the most recent in violet. Curves are ordered chronologically according to rainbow colors.}\label{fig:1}
\end{figure}

To understand the main features of the data, Figure~\ref{fig:1} presents rainbow plots of age- and sex-specific life-table deaths in Australia from 1921 to 2020, grouped by single year. The time ordering of the curves follows the color code of a rainbow, where data from the distant past are shown in red, and the more recent data are shown in purple \citep[see, e.g.,][]{HU07, HS09}. Both figures demonstrate a decreasing trend in infant deaths and a typical negatively skewed distribution for the life-table deaths, where the peaks shift to higher ages for both females and males. This shift is the primary evidence of longevity risk, which challenges superannuation, insurers, and pension funds, especially in the selling and risk management of annuity products \citep[see, e.g.,][]{DDG07, DHR11}.

\begin{figure}[!htb]
\centering
\includegraphics[width=8.42cm]{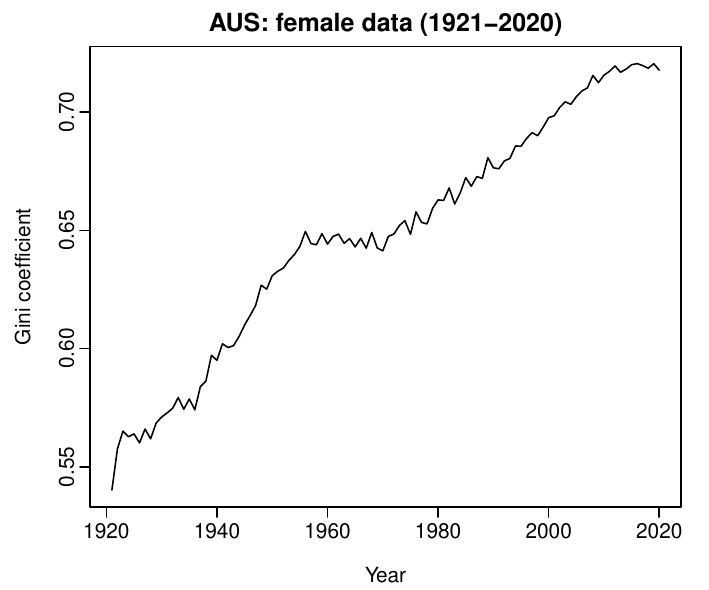}
\quad
\includegraphics[width=8.42cm]{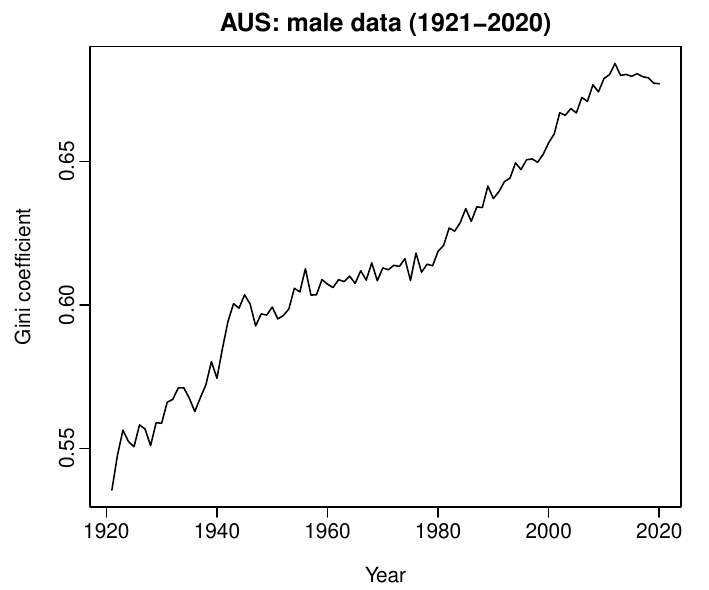}
\caption{\small Gini coefficients for Australian age- and sex-specific life-table deaths from 1921 to 2020. When the Gini coefficient is close to zero, it indicates perfect inequality across ages in the life-table deaths. When the Gini coefficient is close to one, it indicates perfect equality when death occurs at the same age.}\label{fig:2}
\end{figure}

The spread of the distribution indicates variability in lifespan. A decrease in variability over time can be directly observed and measured by the interquartile range or Gini coefficient of the life-table deaths \citep[see, e.g.,][]{WH99}. In economics, the Gini coefficient summarizes the degree of income equality with a single value, ranging from 0 (perfect equality) to 1 (perfect inequality). Let $\mathsf{p}$ be the fraction of the population that holds $L(\mathsf{p})$ proportion of the whole income. It can be expressed as: $G = 2\int^1_0 [\mathsf{p} - L(\mathsf{p})]d\mathsf{p}$. As income and deaths are inversely related, a value of zero indicates perfect inequality among ages in life-table deaths, and a value of 1 indicates ultimate equality, implying that deaths occur at the same age. Using the Australian data, Figure~\ref{fig:2} presents a time series of Gini coefficients, where the life-table deaths provide essential insights into longevity and lifespan variability that cannot be grasped directly from examining the trends in mortality rates or survival functions.

\section{CoDa with the dynamic factor model}\label{sec:3}

\subsection{clr transformation}\label{sec:3.0}

Let $d_t(u)$ be the age-specific life-table death count for age $u$ at year $t$. For a given year $t$, compositional data are defined as a random vector of $p$ non-negative components, the sum of which is a specified constant. Between the non-negativity and summability constraints, the sample space of functional compositional data is a simplex. 

When age $u$ is treated as a continuum, the clr transformation can be written as
\begin{equation*}
\text{clr}[d_t(u)] := \X_{t}(u) = \ln d_t(u) - \frac{1}{\eta}\int_{u_1}^{u_D} \ln d_{t}(u)du, \qquad t=1,2,\dots,n,
\end{equation*}
where $\eta = u_D - u_1$ denotes the length of the age interval, $u_D$ denotes the last age group, and $\frac{1}{\eta}\int_{u_1}^{u_p} \ln d_t(u)du$ is the geometric mean. Via the inverse clr transformation, the forecast life-table deaths can be obtained as
\begin{equation*}
\widehat{d}_{n+h|n}(u) = \frac{\exp[\widehat{\X}_{n+h|n}(u)]}{\int^{u_D}_{u_1} \exp[\widehat{\X}_{n+h|n}(u)]du}\times 10^5,
\end{equation*}
where $10^5$ is the life-table radix.

\subsection{Dynamic factor model}\label{sec:3.1}

In the functional time series literature, recent advances in modeling and forecasting have focused on stationary series, and extending these methods to nonstationary series is challenging. Without stationarity, certain statistics, such as the long-run covariance function, are not well-defined. Age-specific mortality data often exhibit nonstationarity, which can be confirmed by a functional KPSS test of \cite{KY16} and \cite{CP19}. Using the fkpsst function in the \textit{STFTS} package \citep{CP21} in \Rlogo, we compute the $p$-value under the null hypothesis of trend stationarity. For both data, the $p$-value is less than $2.2 \times 10^{-16}$, indicating a strong nonstationarity.

A solution is to take first-order differencing of the original functional time series to achieve stationarity \citep[see, e.g.,][]{MGG22, SY25}. The idea of successive differencing has also been considered in the estimation of the long-memory parameter in nonstationary functional time series \citep[see, e.g.,][]{LRS23}. After taking the first-order differencing, we apply the functional KPSS test again and obtain the $p$-value of one, indicating stationarity. With the differenced series, we use a kernel sandwich method to estimate the long-run covariance function from which we extract functional principal components. By projecting these components onto the original functional time series, we obtain the primary principal component scores. 

With the principal component decomposition, we also obtain its model residual functions, in which an independent test, as described in \cite{GK07}, can be applied to examine whether any temporal patterns remain. If the residuals still exhibit temporal dependence, we perform a second-stage functional principal component decomposition to extract the principal components of the residuals. By projecting the residuals onto the principal components of the residuals, we obtain the principal component scores of the residuals. Conditional on the observed data, the estimated mean function, and functional principal components, the interval forecasts can be obtained by forecasting the primary and residuals' scores.

For a nonstationary functional time series, a characteristic is shown through at least one of the principal component scores $(\widehat{\beta}_{t,1},\dots,\widehat{\beta}_{t,\widehat{N}})$, where $\widehat{N}$ is the retained number of components. We assume that the nonstationary principal component scores exhibit $I(1)$ scalar-valued processes. Let $r\in \{1,2,\dots,\widehat{N}\}$ be the number of principal component scores that is $I(1)$ processes. Although it is possible to have $r=\widehat{N}$, it is common to observe $r<\widehat{N}$ and the remaining $(\widehat{N} - r)$ processes are stationary \citep[see, e.g.,][]{CKP16}. In the latter case, a part of the underlying process resides in the nonstationary space, while the remaining part resides in the stationary space \citep{NSS23, SS24}.

To estimate the two-stage functional principal components, we consider the following procedure:
\begin{enumerate}
\item[1)] Compute the estimated long-run covariance function based on the first-order differencing of the functional time series, denoted by $\W_s(u) = \Delta \X_{s}(u) = \X_{s}(u) - \X_{s-1}(u)$, for $s=2,\dots,n$. To compute the long-run covariance function, we resort to a kernel sandwich estimator of \cite{RS17} with a plug-in bandwidth. Given its observations $\W_2(u),\dots,\W_n(u)$, a natural estimator of the long-run covariance is
\begin{equation*}
\widehat{C}_{h,q}(u,v) = \sum^{\infty}_{\ell = -\infty}W_q(\frac{\ell}{h})\widehat{\gamma}_{\ell}(u,v),
\end{equation*}
where $\ell$ denotes a lag variable, $h$ denotes a bandwidth selected by a plug-in algorithm of \cite{RS17}, $\widehat{\gamma}_{\ell}(u,v)$ is an empirical estimator of the autocovariance function at lag $\ell$, and $W_q(\cdot)$ is the $q$\textsuperscript{th} order kernel function. The popular first and second-order kernel functions include Bartlett, Parzen, Tukey-Hanning, and quadratic spectral kernel functions. Here, we consider the Bartlett kernel function as $q=1$.
\item[2)] From the estimated long-run covariance function, we compute the estimated functional principal components $\{\widehat{\zeta}_k\}$ and their associated scores $\widehat{\beta}_{t,k} = \langle \X_t(u), \widehat{\zeta}_k(u)\rangle$ for $k=1,2,\dots,r$, where $\langle \cdot \rangle$ denotes the $L^{2}$ inner product. The value of $r$ can be set as one in \cite{LC92} or six in \cite{HBY13}.
\item[3)] Compute the functional residuals: $\Z_t(u) = \X_t(u) - \overline{\X}(u) - \sum_{k=1}^{r}\widehat{\beta}_{t,k}\widehat{\zeta}_k(u)$.
\item[4)] Apply the independence test of \cite{GK07} to $\bm{\Z}(u) = [\Z_1(u),\dots,\Z_n(u)]$. If $\bm{\Z}(u)$ is independent, terminate; Else, compute the estimated long-run covariance function of~$\bm{\Z}(u)$.
\item[5)] Let $(\widehat{N}-r)$ be the estimated number of principal components, which can be set as one in \cite{LC92} or six as in \cite{HBY13}.
\item[6)] Obtain all estimated functional principal components $\{\widehat{\zeta}_{\omega}(u), \omega = r+1, r+2, \dots, \widehat{N}\}$ and their associated principal component scores $\widehat{\beta}_{t,\omega} = \langle \Z_t(u),\widehat{\zeta}_{\omega}(u)\rangle$.
\end{enumerate}

\subsection{Construction of prediction intervals via nonparametric bootstrapping}\label{sec:3.2}

We consider two sources of uncertainty: truncation errors in the functional principal component decomposition and forecast errors in the forecast principal component scores. Since the principal component scores are regarded as surrogates of the original functional time series, these principal component scores capture the temporal dependence structure inherited in the original functional time series \citep[see, e.g.,][]{Paparoditis18, Shang18}. Conditional on the estimated mean function and functional principal components, we can generate bootstrap forecasts $\bm{\X}^*_{n+h}$ by bootstrapping the forecast principal component scores and residual functions.

To model temporal dependence, we resort to a univariate time-series method. For $h=1,2,\dots,20$, we can obtain multi-step-ahead forecasts for the principal component scores, $\{\widehat{\beta}_{1,k},\dots,\widehat{\beta}_{n,k}\}$ for $k=1, 2, \dots,r$. For each $k$, let the $h$-step-ahead forecast errors be given by
\begin{equation*}
\widehat{\sigma}_{t,h,k} = \widehat{\beta}_{t,k} - \widehat{\beta}_{t|t-h,k},\qquad t=h+1,\dots,n.
\end{equation*}
These errors can be sampled with replacement to give bootstrap samples of $\beta_{n+h,k}$,
\begin{equation*}
\widehat{\beta}_{n+h|n,k}^{(b)} = \widehat{\beta}_{n+h|n,k}+\widetilde{\sigma}_{*,h,k}^{(b)}, \qquad b=1, 2, \dots,B,
\end{equation*}
where $B=1,000$ symbolizes the number of bootstrap replications and $\widetilde{\sigma}_{*,h,k}^{(b)}$ are sampled with replacement from $\{\widehat{\sigma}_{h+1,h,k},\dots,\widehat{\sigma}_{n,h,k}\}$.

Analogously, we can also obtain multi-step-ahead forecasts for the principal component scores, $\{\widehat{\beta}_{1,\omega},\dots,\widehat{\beta}_{n,\omega}\}$ for $\omega = 1,\dots,(\widehat{N}-r)$. Denote the $h$-step-ahead forecast errors as
\begin{equation*}
\nu_{t,h,\omega} = \widehat{\beta}_{t,\omega} - \widehat{\beta}_{t|t-h,\omega}.
\end{equation*}
These errors can then be sampled with replacement to give bootstrap samples of $\beta_{n+h,\omega}$
\begin{equation*}
\widetilde{\beta}_{n+h|n,\omega}^{(b)} = \widehat{\beta}_{n+h|n,\omega}+ \widetilde{\nu}_{*,h,\omega}^{(b)},
\end{equation*}
where $\widetilde{\nu}_{*,h,\omega}^{(b)}$ are sampled with replacement from $\{\widehat{\nu}_{h+1,h,\omega},\dots,\widehat{\nu}_{n,h,\omega}\}$.

Via the two-step functional principal component analyses, the functional time series $\X_t(u)$ can be approximated by
\begin{equation*}
\X_t(u) = \overline{\X}(u) + \sum_{k=1}^{r}\widehat{\beta}_{t,k}\widehat{\zeta}_k(u) + \sum_{\omega=1}^{\widehat{N}-r}\widehat{\beta}_{t,\omega}\widehat{\zeta}_{\omega}(u) + \Y_t(u),
\end{equation*}
where $\Y_t(u)$ denotes a model residual function for time $t$. In extreme events like COVID-19, it is expected the residual function is larger than those in the standard period. To capture its uncertainty, we can bootstrap the residual functions by sampling with replacement from $\{\Y_1(u),\Y_2(u), \dots,\Y_n(u)\}$.

Adding all sources of variability, we obtain $B$ variants for $\X_{n+h}(u)$:
\begin{equation*}
\X_{n+h|n}^{(b)}(u) = \overline{\X}(u) + \sum_{k=1}^{r}\widetilde{\beta}_{n+h|n,k}^{(b)}\widehat{\zeta}_{k}(u)+\sum_{\omega=1}^{\widehat{N}-r}\widetilde{\beta}_{n+h|n,\omega}^{(b)}\widehat{\zeta}_{\omega}(u)+\Y_{*}^{(b)}(u).
\end{equation*}
With the bootstrapped $\X_{n+h|n}^{(b)}(u)$, we take the inverse clr transformation to obtain $\widehat{d}_{n+h|n}^{(b)}(u)$. The pointwise prediction intervals of the life-table deaths are obtained by taking $\alpha/2$ and $(1-\alpha/2)$ quantiles at the $100(1-\alpha)\%$ nominal coverage probability, where $\alpha$ denotes a level of significance, customarily $\alpha = 0.2$ or 0.05.

\subsection{CoDa Lee-Carter (CoDa-LC) bootstrap method}\label{sec:3.3}

The construction of prediction intervals for CoDa has previously been considered by \cite{BCO+17}, which was built on the \citeauthor{KP06}'s \citeyearpar{KP06} model. It allows one to consider two sources of forecast uncertainty: 
\begin{inparaenum}
\item[1)] estimates of the parameters; and 
\item[2)] extrapolated values of the time index.
\end{inparaenum}
The following steps apply to the prediction intervals of the CoDa-LC model.
\begin{enumerate}
\item[1)] After applying the clr transformation, \cite{BCO+17} apply principal component analysis to obtain the first (few) estimated principal component and its associated scores, as well as the residual matrix.
\item[2)] Bootstrapped residuals can be obtained by sampling with replacement from the original residual matrix. 
\item[3)] With the bootstrap residuals, they then add them to the estimated mean term to form the bootstrapped data samples.
\item[4)] With each replication of the bootstrap data samples, one could then re-apply singular value decomposition to obtain the bootstrapped principal components and their associated scores.
\item[5)] The bootstrapped principal component scores can be extrapolated using a univariate time-series forecasting method, such as the exponential smoothing of \cite{HKO+08}.
\item[6)] By multiplying the bootstrapped forecasts of the principal component scores with the estimated principal components and mean function, the bootstrap forecasts of the life-table deaths can be obtained via back-transformation.
\end{enumerate}

\section{Comparison of interval forecast accuracy}\label{sec:4}

\subsection{Expanding-window approach}\label{sec:4.1}

An expanding window analysis of a time-series model is employed to evaluate model and parameter stability over time and predict accuracy across various horizons. The expanding-window analysis determines the constancy of a model's parameter by computing parameter estimates and their resultant forecasts over an expanding window of a fixed size through the sample \citep[for details][pp. 313-314]{ZW06}. For instance, using the first 80 observations from 1921 to 2000 in the Australian life-table deaths, we produce 1- to 20-step-ahead interval forecasts. Through an expanding-window approach, we re-estimate the parameters in the time-series forecasting models using the first 81 observations from 1921 to 2001. Forecasts from the estimated models are then produced for 1- to 19-step-ahead interval forecasts. We iterate this process by increasing the sample size by one year until reaching the end of the data period in 2020. This process produces 20 one-step-ahead forecasts, 19 two-step-ahead forecasts, $\dots$, and one 20-step-ahead forecast. We compare these forecasts with the holdout samples to determine the accuracy of the out-of-sample interval forecast. In Figure~\ref{fig:3}, we present a diagram of the expanding-window approach:
\tikzset{decorate with/.style={fill=cyan!20,draw=cyan}}
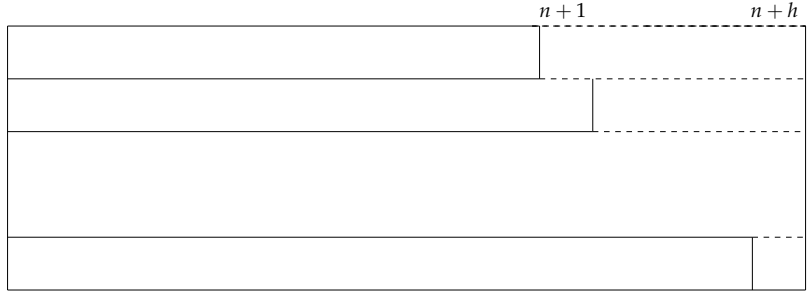
\begin{figure}[!htb]
\begin{center}
\scalebox{.7}{
\begin{tikzpicture}
\draw (0,0) -- (0, 5); 
\draw (0,0) -- (15, 0); 
\draw (10, 5) -- (10, 4);
\draw (0,4) -- (10, 4);
\draw[dashed] (10, 4) -- (15, 4);
\draw (0,3) -- (11, 3);
\draw (11, 3) -- (11, 4);
\draw[dashed] (11, 3) -- (15, 3);
\draw (0,5) -- (10, 5);
\draw[dashed] (12, 5) -- (15, 5);
\draw (0,1) -- (14, 1);
\draw (14, 1) -- (14, 0);
\draw[dashed] (14, 1) -- (15, 1);
\draw (15, 0) -- (15, 5);
\draw (9.86, 5)[dashed] node[anchor=south west]{$n+1$} -- (15, 5);
\draw (15, 5)[dashed] node[anchor=south east]{$n+h$} -- (9.86, 5);
\end{tikzpicture}}
\end{center}
\caption{\small A sketch of the expanding-window forecasting scheme for various horizons $h=1,2,\dots, 20$.}\label{fig:3}
\end{figure}

\subsection{Interval forecast evaluation}\label{sec:4.2}

The evaluation of interval forecast accuracy is implemented and compared as follows: 
\begin{inparaenum}
\item[1)] Comparison between the CoDa-DFM and the CoDa-LC bootstrap methods;
\item[2)] Comparison between the different numbers of retained components.
\end{inparaenum}
For each year in the forecasting period, the one- to 20-step-ahead prediction intervals were calculated at the customarily 80\% and 95\% nominal coverage probabilities and were then tested against the empirical coverage probabilities \citep[see also][]{SB94, TSL07, SBH11}. The empirical coverage probability is the proportion of the holdout samples that fall into the computed prediction intervals. It can be expressed as
\begin{equation*}
\text{ECP}(h) = 1 - \frac{1}{(21-h)\times D}\sum^{21-h}_{\zeta=1}\sum^D_{i=1}\left\{\mathds{1}[d_{n+\zeta}(u_i)>\widehat{d}^{\text{ub}}_{n+\zeta|n}(u_i)] + \mathds{1}[d_{n+\zeta}(u_i)<\widehat{d}^{\text{lb}}_{n+\zeta|n}(u_i)]\right\},
\end{equation*}
where $\zeta = 1, 2, \dots,(21-h)$ denotes the number of observations in the forecasting period for a given horizon~$h$, $D$ denotes the number of ages, $\widehat{d}^{\text{ub}}_{n+h|n}(u_i)$ and $\widehat{d}^{\text{lb}}_{n+h|n}(u_i)$ denote the upper and lower bounds of the prediction intervals, and $\mathds{1}[\cdot]$ represents the binary indicator function. The coverage probability difference (CPD) is defined as
\begin{equation*}
\text{CPD}(h) = \left|\text{Empirical coverage}(h) - \text{Nominal coverage}\right|. 
\end{equation*}
The lower the CPD$(h)$ value, the better the forecasting uncertainty is modeled. Averaged over 20 different forecast horizons, we compute the averaged $\overline{\text{ECP}}$ and $\overline{\text{CPD}}$, defined as
\begin{align*}
\overline{\text{ECP}} &= \frac{1}{20}\sum^{20}_{h=1}\text{ECP}(h), \\
\overline{\text{CPD}} &= \frac{1}{20}\sum^{20}_{h=1}\text{CPD}(h).
\end{align*}

\subsection{Results of the Australian and UK life-table deaths}\label{sec:4.3}

\begin{figure}[!htb]
\centering
\includegraphics[width=8.42cm]{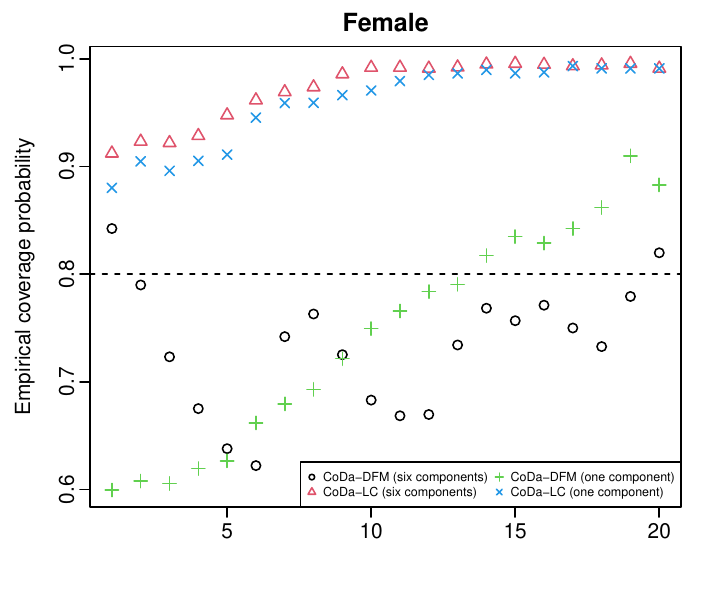}
\quad
\includegraphics[width=8.42cm]{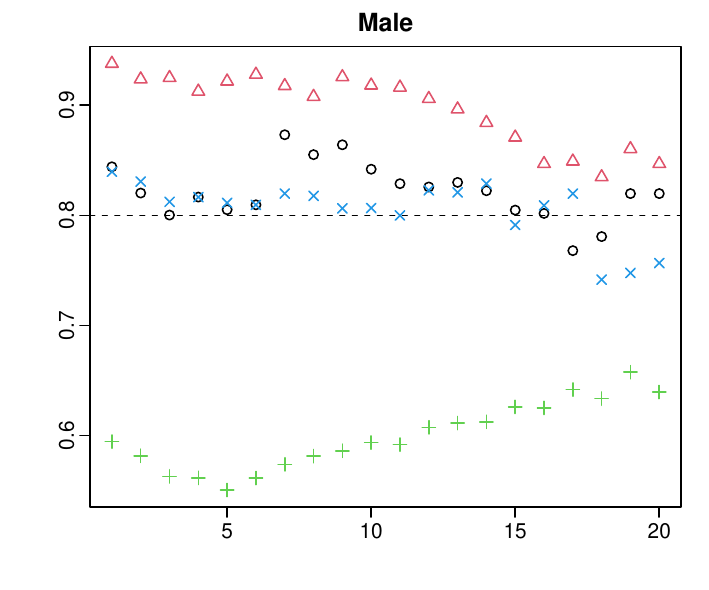}
\\
\includegraphics[width=8.42cm]{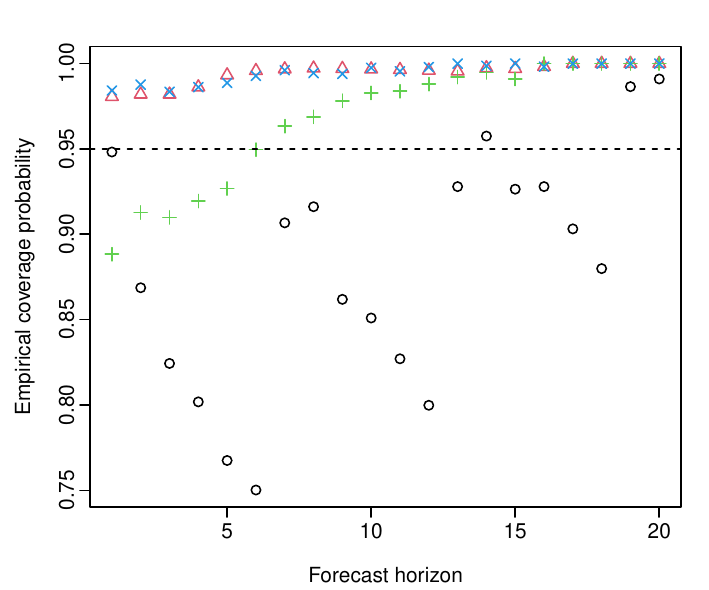}
\quad
\includegraphics[width=8.42cm]{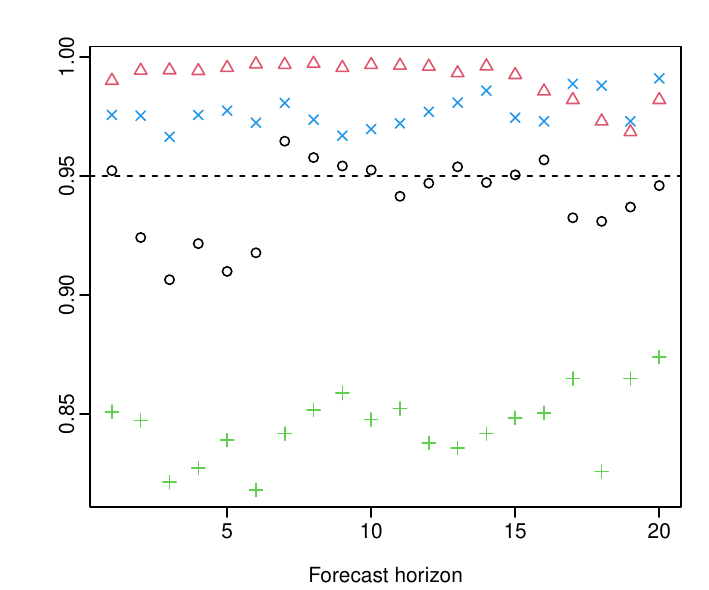}
\caption{\small At the nominal coverage probabilities of 80\% and 95\%, we compute the empirical coverage probabilities of the pointwise prediction intervals from the CoDa-DFM and CoDa-LC bootstrap methods for 20 different forecast horizons based on female and male life-table deaths in Australia from 2001 to 2020.}\label{fig:4}
\end{figure}

We present the interval forecast results for the two datasets: Australia and the UK. At nominal coverage probabilities of 80\% and 95\%, we compare the CoDa-DFM and CoDa-LC bootstrap methods, as well as two methods for selecting the number of retained components. From Figure~\ref{fig:4}, the CoDa-LC bootstrap method produces excessively higher empirical coverage probabilities for the female data, reflecting much wider prediction intervals. 

The CoDa-DFM bootstrap method, with six components, produces accurate one-step-ahead interval forecasts; however, the interval forecast accuracy fluctuates as the forecast horizon increases. With only one component, the CoDa-DFM bootstrap method underestimates the nominal coverage probability at shorter horizons and overestimates it at longer horizons. For the male data, using six components when implementing the CoDa-DFM method is advantageous.

\begin{figure}[!htb]
\centering
\includegraphics[width=8.42cm]{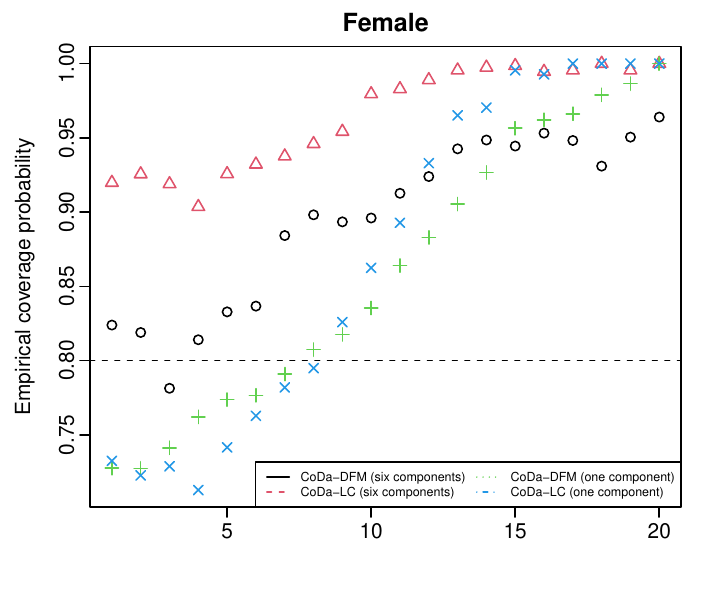}
\quad
\includegraphics[width=8.42cm]{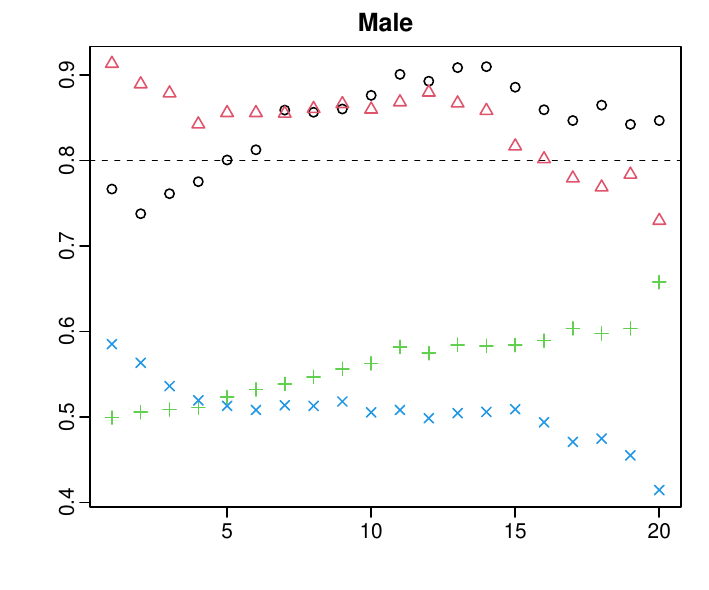}
\\
\includegraphics[width=8.42cm]{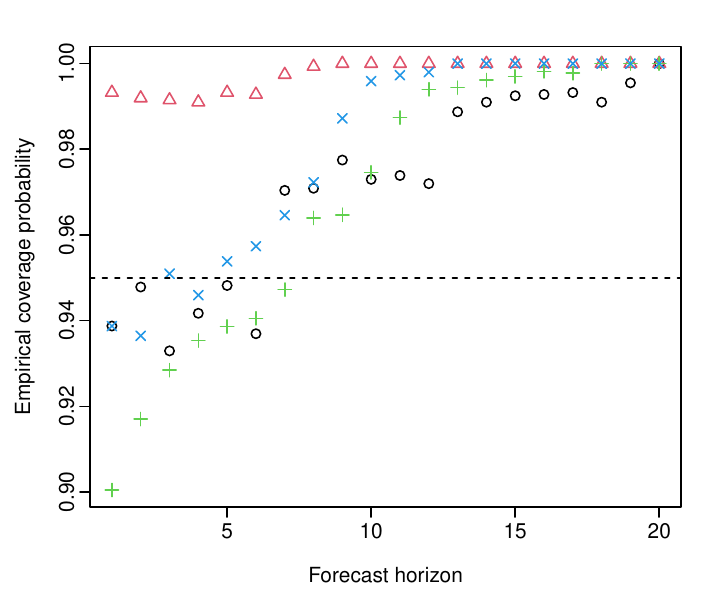}
\quad
\includegraphics[width=8.42cm]{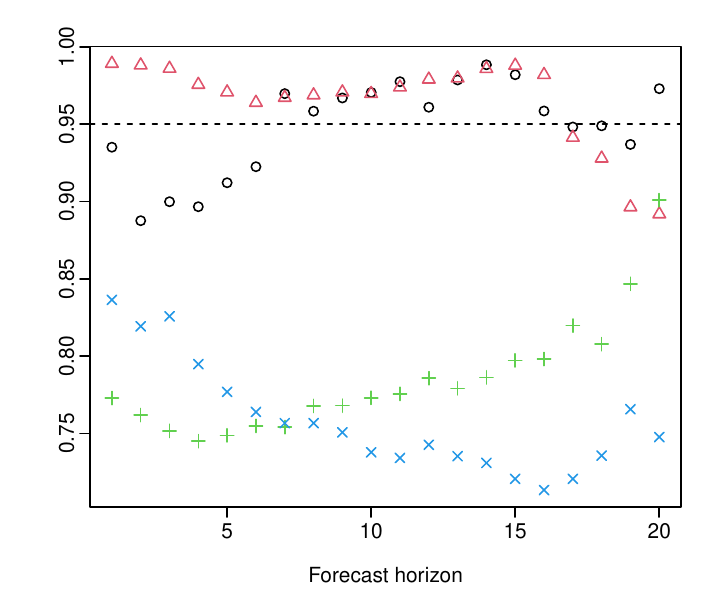}
\caption{\small At the nominal coverage probabilities of 80\% and 95\%, we compute the empirical coverage probabilities of the pointwise prediction intervals from the CoDa-DFM and CoDa-LC bootstrap methods for 20 different forecast horizons using the female and male life-table deaths in the UK from 2002 to 2021.}\label{fig:5}
\end{figure}

Using the life-table deaths in the UK, the CoDa-LC bootstrap method produces higher empirical coverage probabilities for the female data in Figure~\ref{fig:5}. The CoDa-DFM bootstrap method, with six components, yields the most accurate one-step-ahead interval forecasts. Still, the empirical coverage probability tends to exceed the nominal coverage probability as the forecast horizon increases. There is severe underfitting for the UK male data when the number of components is kept at one.

By taking the simple average over 20 forecast horizons, we compare the interval forecast accuracy, as measured by the empirical coverage probability and the coverage probability difference, between the CoDa-DFM and CoDa-LC bootstrap methods. For selecting the number of retained components, we set $r = (\widehat{N}-r) = 6$ or one. In Table~\ref{tab:1}, the CoDa-DFM with six components is recommended for achieving a nominal coverage probability of 80\% in modeling female life-table deaths. At the nominal coverage probability of 95\%, the CoDa-DFM with one component is advantageous. For modeling the male life-table deaths, the CoDa-DFM with six components is recommended.

\begin{table}[!htb]
\tabcolsep 0.095in
\centering
\caption{\small We calculate the empirical coverage probabilities and coverage probability deviances of the CoDa-DFM and CoDa-LC bootstrap methods at nominal coverage probabilities of 80\% and 95\% for the Australian and UK life-table deaths. We also consider two ways of selecting the number of components.}\label{tab:1}
\begin{tabular}{@{}lllccccccccc@{}}
\toprule
	&	& 	& \multicolumn{4}{c}{$r = 6$, $\widehat{N}-r=6$} & & \multicolumn{4}{c}{$r = 1$, $\widehat{N}-r=1$} \\
\cmidrule{4-7}\cmidrule{9-12}
		& &	& \multicolumn{2}{c}{CoDa-DFM} & \multicolumn{2}{c}{CoDa-LC} & & \multicolumn{2}{c}{CoDa-DFM} & \multicolumn{2}{c}{CoDa-LC} \\	
Country &  Nominal & Sex & ECP & CPD & ECP & CPD & & ECP & CPD & ECP & CPD \\\midrule 
AUS & 80\% 	&  F & 0.7327 & \textBF{0.0735} & 0.9724 & 0.1724 & & 0.7441 & 0.0937 & 0.9589 & 0.1589 \\ 
	& 		&  M & 0.8216 & \textBF{0.0268} & 0.8965 & 0.0965 & & 0.5998 & 0.2002 & 0.8055 & 0.0217 \\ 
\\
	& 95\% 	&  F & 0.8812 & 0.0773 & 0.9944 & 0.0444 & & 0.9675 & \textBF{0.0368} & 0.9948 & 0.0448 \\ 
	& 		&  M & 0.9402 & \textBF{0.0140} & 0.9909 & 0.0409 & & 0.8450 & 0.1050 & 0.9769 & 0.0269 \\ 
\\
UK 	& 80\% 	& F & 0.8949 & 0.0968 & 0.9646 & 0.1646 & & 0.8594 & \textBF{0.0895} & 0.8708 & 0.1130 \\ 
  	& 		& M & 0.8433 & 0.0592 & 0.8416 & \textBF{0.0555} & & 0.5622 & 0.2378 & 0.5054 & 0.2946 \\ 
\\
	& 95\% 	&  F & 0.9714 & \textBF{0.0268} & 0.9975 & 0.0475 & & 0.9688 & 0.0330 & 0.9799 & 0.0328 \\ 
	& 		&  M & 0.9486 & \textBF{0.0248} & 0.9648 & 0.0290 & & 0.7848 & 0.1652 & 0.7584 & 0.1916 \\
\bottomrule
\end{tabular}
\end{table}

\section{Conclusion}\label{sec:5}

We present a nonparametric bootstrap procedure based on the dynamic factor model within the CoDa framework. Our findings are listed as follows: 
\begin{inparaenum}
\item[1)] We show its superior finite-sample performance when retaining the first six principal components for the male data from Australia and the UK.
\item[2)] For the comparably less volatile female data, the bootstrap method with the first principal component seems sufficient to model forecast uncertainty.
\item[3)] Particularly for the female data, the CoDa-LC bootstrap method tends to overestimate the nominal coverage probabilities, leading to much wider prediction intervals.
\end{inparaenum}

There are at least two ways in which the methodology presented can be further extended:
\begin{inparaenum}
\item[1)] The transformed data may exhibit a long-memory feature, where a fractional-order differencing may help model the non-stationary series.
\item[2)] We consider the life-table deaths at the national level for Australia and the UK; it may be interesting to investigate the interval forecast accuracy for subnational life-table deaths with different geographical regions.
\end{inparaenum}



\newpage
\bibliographystyle{agsm}
\bibliography{CoDa_sieve}

\end{document}